# Nanospot Angle-Resolved Photoemission Study of Bernal-Stacked Bilayer Graphene on Hexagonal Boron Nitride: Band Structure and Local Variation of Lattice Alignment

Frédéric Joucken,[1] Eberth A. Quezada-López,[1] Jose Avila,[2] Chaoyu Chen,[2] John L. Davenport,[1] Hechin Chen,[1] Kenji Watanabe,[3] Takashi Taniguchi,[3] Maria Carmen Asensio,[4*] Jairo Velasco Jr.[1*]

[1]Department of Physics, University of California, Santa Cruz, California 95060, USA

[2] ANTARES Beamline, Synchrotron SOLEIL & Université Paris-Saclay, L'Orme des Merisiers, Saint Aubin-BP 48, 91192 Gif sur Yvette Cedex, France

[3] National Institute for Materials Science, 1-1 Namiki, Tsukuba, 305-0044, Japan

[4] Madrid Institute of Materials Science (ICMM), Spanish Scientific Research Council (CSIC), Cantoblanco, E-28049 Madrid, Spain

*  Correspondence to: mc.asensio@csic.com, jvelasc5@ucsc.edu

**Abstract:**

**Hexagonal boron nitride ($h$BN) is the supporting substrate of choice for two-dimensional material devices because it is atomically flat and chemically inert. However, due to the small size of mechanically exfoliated $h$BN flakes, electronic structure studies of 2D materials supported by $h$BN using angle-resolved photoemission spectroscopy (ARPES) are challenging. Here we investigate the electronic band structure of a Bernal-stacked bilayer graphene sheet on a hexagonal boron nitride (BLG/$h$BN) flake using nanospot ARPES (nanoARPES). By fitting high-resolution energy *vs.* momentum electronic band spectra, we extract the tight-binding parameters for BLG on $h$BN. In addition, we reveal spatial variations of the alignment angle between BLG and $h$BN lattices *via* inhomogeneity of the electronic bands near the Fermi level. We confirmed these findings by scanning tunneling microscopy measurements obtained on the same device. Our results from spatially resolved nanoARPES measurements of BLG/$h$BN heterostructures are instrumental for understanding experiments that utilize spatially averaging techniques such as electronic transport and optical spectroscopy.**

**Keywords:** Bilayer graphene, graphene, $h$BN, van der Waals heterostructure, NanoARPES, electronic structure, tight-binding, ARPES, STM

**Main text:**

The recent development of nanospot angle-resolved photoemission spectroscopy (nanoARPES)[1–4] allows unprecedented spatial mapping of the electronic band structure of two-dimensional material (2D) heterostructures.[5–9] In particular, by applying this spatially-resolved technique to heterostructures that use hexagonal boron nitride (hBN) as a supporting substrate, reliable, direct, and insightful visualization of top lying 2D material band structures can be achieved.[5–9] Recent spatially-resolved ARPES studies on monolayer graphene/hBN and $WS_2$/hBN have carefully examined the band structure of the top lying 2D material and found evidence for polarons[6] and band gap renormalization.[8] Bernal-stacked bilayer graphene (BLG) that is supported on hBN is also of fundamental interest because of the peculiar physics that has been recently revealed in this system, such as excitons with pseudospin texture,[10] and indications of non-Abelian excitations.[11] However, direct spatially-resolved electronic structure investigation of BLG/hBN heterostructures with ARPES is lacking.

We present in this Rapid Communication a direct spatial mapping of the electronic structure of a BLG/hBN heterostructure using nanoARPES in combination with scanning tunneling microscopy (STM) imaging. We extract tight-binding parameters from high resolution ARPES spectra; thus, allowing a comparison between BLG/hBN parameters with those from previous measurements on BLG resting on different substrates and acquired *via* different techniques. In addition, we show that direct access to the band structure with submicron spatial resolution offered by nanoARPES reveals small spatial variations of lattice alignment between BLG and the supporting hBN. Our results provide important fundamental insight on band structure parameters and nanoscale alignment between 2D materials. This insight can be used for improved modeling and further understanding of the interesting physics hosted in BLG/hBN heterostructures.

The sample was fabricated using the method reported by Zomer et al.[12] with stencil mask-aided metallization for electrodes. We used high purity $h$BN crystals synthesized by Taniguchi et al.,[13] exfoliated to a 50-nm thickness and deposited on $SiO_2$/Si chip with an oxide thickness of 285 nm. BLG is exfoliated from graphite and deposited onto methyl methacrylate polymer and transferred onto $h$BN sitting on the $SiO_2$/Si chip.

Our study was composed of two different characterization techniques (nanoARPES and STM) that was applied to our BLG/$h$BN heterostructure. The NanoARPES experiments were carried out at the ANTARES beamline of synchrotron SOLEIL. It is equipped with a Fresnel zone plate (FZP) to focalize the beam and an order selection aperture to eliminate higher diffraction orders. The sample was mounted on a nano-positioning stage which was placed at the coincident focus point of the electron analyzer and the FZP. The photoelectron spectra were obtained using a hemispherical analyzer (MBS A1) equipped with electrostatic lenses allowing to perform Fermi surface measurements without rotating the sample. All photoemission measurements were performed at a temperature of ~ 100 K, at a photon energy of 100 eV, and with an overall energy and momentum resolution better than 35 meV and 0.01 Å$^{-1}$, respectively. The sample was annealed at ~300 °C for two hours before the measurements. The STM measurements were conducted in ultra-high vacuum (UHV) at a pressure below 2 x 10$^{-10}$ mbar and at a temperature of 4.8K in a Createc STM. The bias was applied to the sample with respect to the tip. The tips were electrochemically-etched tungsten tips and were calibrated against the Shockley surface state of Au(111) prior to measurements. The sample was exposed to atmosphere between the two UHV setups. The sample was also annealed at ~300 °C for two hours before the STM measurements.

A dark-field optical image of the device is shown in Fig. 1a, with a schematic in the inset.

The BLG flake (orange) rests onto the hBN flake (blue) lying on the SiO$_2$/Si substrate (pink/grey) and is grounded *via* a gold/chromium electrode (yellow). These components are labeled in the dark-field optical image. A nanoARPES image of the same region is displayed in Fig. 1b. Each pixel level of this image is obtained by integrating the ARPES Energy *vs* momentum (E-k) spectrum obtained at the corresponding position (for details on nanoARPES imaging, *cf.* the discussion below and *e.g.* refs.[3,6,14,15]). There is a clear correspondence between the optical and nanoARPES image. This enables identification of device components in the nanoARPES image.

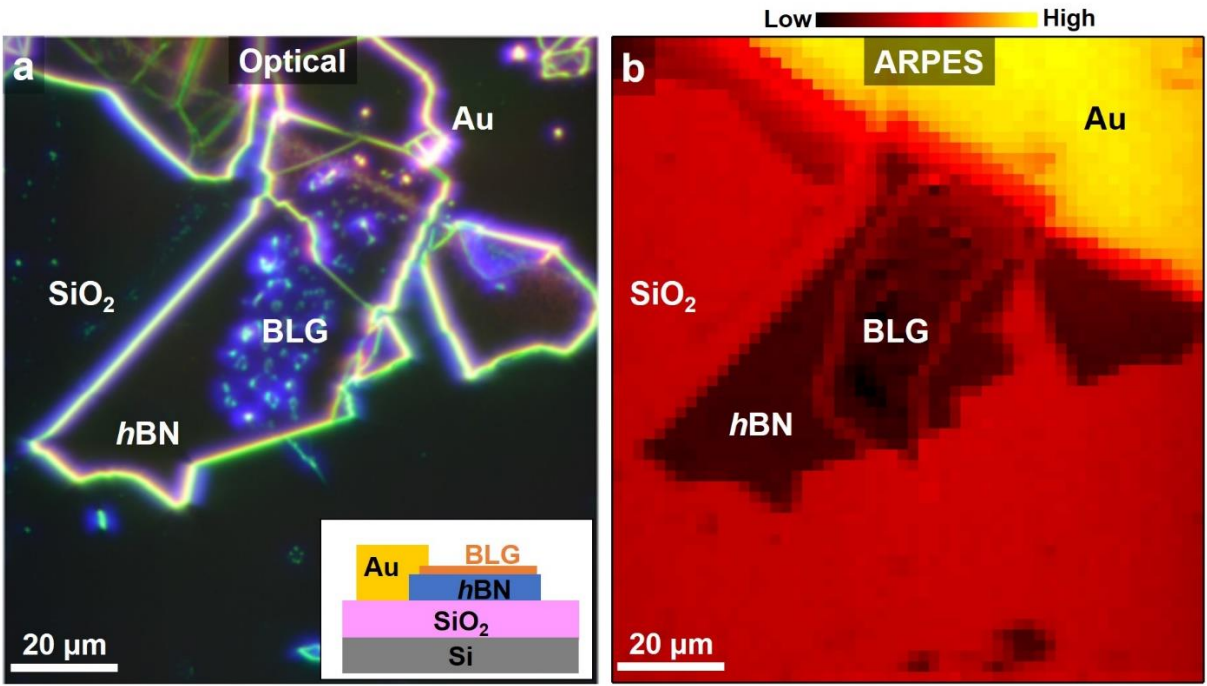

**Figure 1: Optical and photoemission imaging of the heterostructure used for the experiment.**
(a) Optical image of the sample which consists in a Bernal-stacked bilayer graphene/hexagonal boron nitride (BLG/hBN) heterostructure resting upon a SiO$_2$/Si substrate and grounded via a gold electrode. This schematized in the inset with gold electrode in yellow, BLG in orange, hBN in blue, SiO$_2$ in pink, and Si in grey. (b) Photoemission image of the same heterostructure. Each pixel level on the image is given by the integration of an ARPES E-k spectrum at this position. On each image, the components of the device are labeled

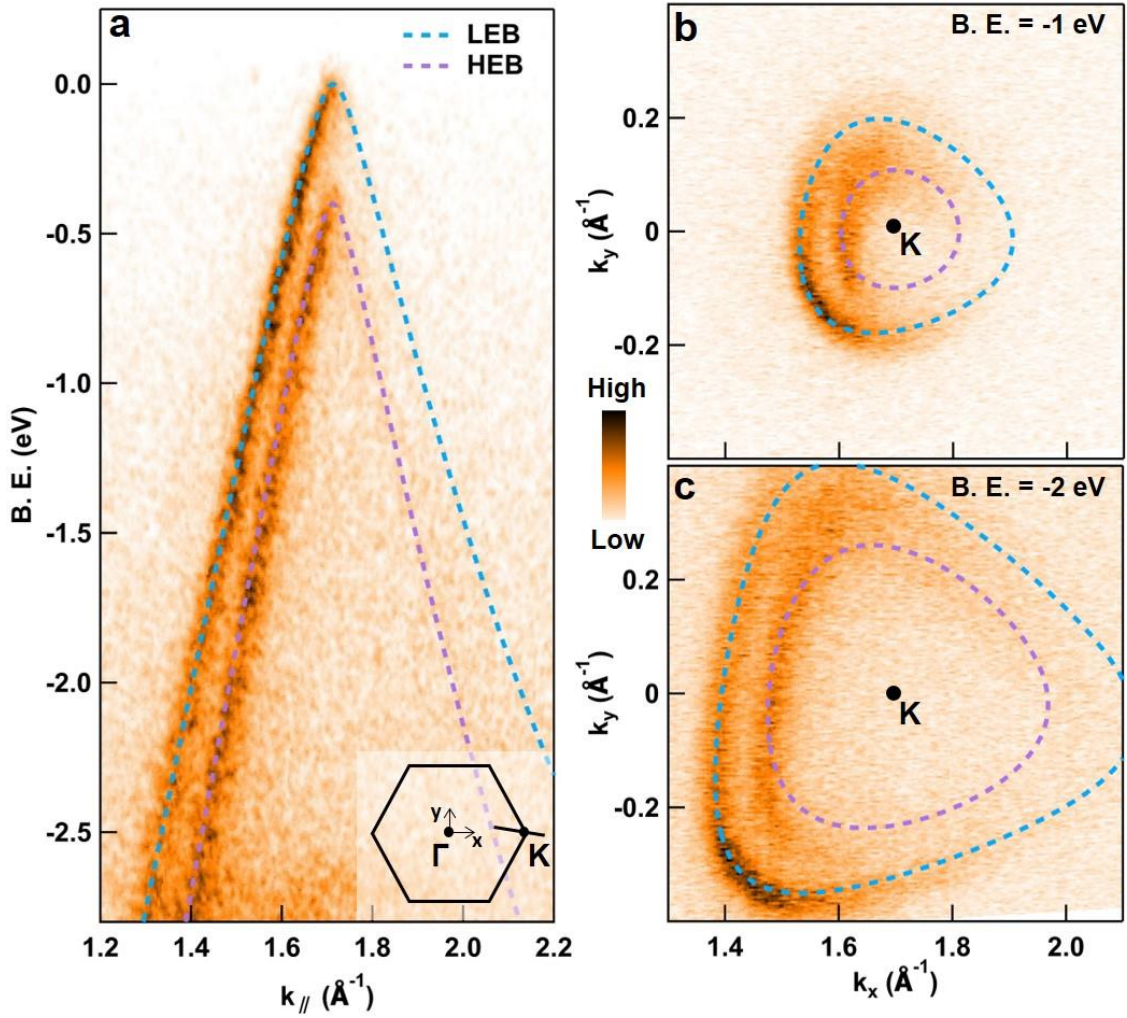

**Figure 2: ARPES characterization and associated tight-binding fits of BLG/hBN.** (a) Experimental ARPES E-k spectrum acquired along the direction indicated in the inset. (b) Experimental ARPES constant energy cut at a binding energy (B. E.) of -1 eV. (c) Experimental ARPES constant energy cut at a B. E. of -2 eV. Each panel contains superimposed best fit tight-binding bands following the conventions McCann and Koshino.[20] Tight-binding parameters used are $\gamma_0$=3.3 eV, $\gamma_1$=0.42 eV, $\gamma_3$=0.07 eV, $\gamma_4$=0.22 eV, and $\Delta'$=0.02 eV.

We first focus on the electronic structure of our BLG flake supported by hBN. Figure 2a displays an ARPES E-k spectrum acquired in the direction indicated in the inset. The measurement direction makes an angle of ~9° with the Γ-K direction. Because of matrix element effects,[16,17] only the part of the band corresponding to k<K (K≃1.70 Å$^{-1}$) is visible. As photoemission only probes occupied states and because the region of the sample being probed is undoped, only the valence bands of BLG are visible in the spectrum.[18,19] We refer to the band closest to the Fermi level as the low energy band (LEB) and the other band as the high energy band (HEB). We show in Figs. 2b and 2c ARPES constant energy cuts acquired around the K point at binding energies (B. E.) of -1 eV and -2 eV, respectively. These constant energy cuts allow another visualization of the matrix element effects.[16,17] We note the three-fold symmetry of the bands around the K point becomes clearly visible at high binding energy (Fig. 2c). This is expected because of the three-fold symmetry of BLG reciprocal space around the K point.

The direct visualization of the bands in conjunction with the quality of our data enables facile quantitative analysis using a tight binding (TB) model to extract band parameters. Fits from such a model are superimposed onto the ARPES E-k spectrum of Fig. 2a and the ARPES constant energy cuts in Figs. 2b and 2c. The conventions we used to compute the TB bands follow McCann and Koshino.[20] Specifically, the TB bands are obtained by solving numerically, at each point in k-space, the following Hamiltonian

$$H = \begin{pmatrix} 0 & -\gamma_0 f(\mathbf{k}) & \gamma_4 f(\mathbf{k}) & -\gamma_3 f^*(\mathbf{k}) \\ -\gamma_0 f^*(\mathbf{k}) & \Delta' & \gamma_1 & \gamma_4 f(\mathbf{k}) \\ \gamma_4 f^*(\mathbf{k}) & \gamma_1 & \Delta' & -\gamma_0 f(\mathbf{k}) \\ -\gamma_3 f(\mathbf{k}) & \gamma_4 f^*(\mathbf{k}) & -\gamma_0 f^*(\mathbf{k}) & 0 \end{pmatrix},$$

with $f(\mathbf{k}) = e^{ik_y a/\sqrt{3}} + 2e^{-ik_y a/2\sqrt{3}} \cos(k_x a/2)$. For undoped non-gated BLG, there are five independent TB parameters to determine: the hopping energies ($\gamma_0$, $\gamma_1$, $\gamma_3$, $\gamma_4$) and $\Delta'$, which is the energy difference between dimer and non-dimer sites.[20] Briefly, the four hopping energies determine the overall velocity of the bands ($\gamma_0$), the energy difference between the tops of the LEB and the HEB ($\gamma_1$), the trigonal warping effect ($\gamma_3$), and the electron-hole asymmetry ($\gamma_4$). Because we do not have access to the unoccupied states with ARPES, we assumed a value of 0.22 eV for $\gamma_4$, as found from STS studies on a similar sample as ours (BLG/$h$BN).[21] Also following previous results, we set Δ' to 0.02 eV (an average of the values found in the literature[22–24]). The remaining parameters were determined by our fits, which yielded: $\gamma_0$=3.3±0.15 eV, $\gamma_1$=0.42±0.05 eV, and $\gamma_3$=0.07±0.1 eV. This corresponds to a band velocity of $v(=\sqrt{3}a\gamma_0/(2\hbar)) = 1.07\pm0.05 \times 10^6$ m/s and an effective mass $m(=\gamma_1/(2v^2)) = 0.032 \pm 0.05 m_e$, where $m_e$ is the bare electron mass.

| Ref. | $\gamma_0$ (eV) | $\gamma_1$ (eV) | $\gamma_3$ (eV) | $\gamma_4$ (eV) | Δ' (eV) | Technique/subtsrate |
|---|---|---|---|---|---|---|
| Malard et al.[29]* | 2.9 | 0.30 | -0.10 | 0.12 | | Raman/SiO$_2$ |
| Ohta et al.[18] | | 0.41-0.46 | 0.12 | | | ARPES/SiC |
| Ohta et al.[30] | 3.24 | 0.48 | | | | ARPES/SiC |
| Zhang et al.[22] | | 0.40 | | 0.15 | 0.018 | IR/SiO$_2$ |
| Lauffer et al.[31] | 3.27 | 0.46 | | | | STM/SiC |
| Henriksen et al.[32] | | 0.43-0.52 | | | | Cycl. Reso./SiO$_2$ |
| Yan et al.[25] | | 0.35 | | | | Raman/SiO$_2$ |
| Kuzmenko et al.[23]* | 3.16 | 0.381 | -0.38 | 0.14 | 0.022 | IR/SiO$_2$ |
| Li et al.[24] | 3.1-3.4 | 0.404 | | 0.16 | 0.018 | IR/SiO$_2$ |
| Zou et al.[26] | 3.43 | | | 0.216 | | Shubnikov–de Haas/SiO$_2$ |
| Mayorov et al.[27] | | | 0.435 | | | Elec. transport/suspended |
| Mallet et al.[38] | 3.7# | 0.38 | | | | STM/SiC |
| Yankowitz et al.[21] | 3.1 | | | 0.22 | | STM/hBN |
| Cheng et al.[19] | -3.21 | 0.61 | 0.39 | 0.15 | | ARPES/SiO$_2$ |
| Lee et al.[28] | 3.1-3.3 | 0.35-0.42 | | 0.06-0.12 | | Elec. transport/hBN |
| This work | 3.3 | 0.42 | 0.07 | | | ARPES/hBN |

**Table 1: Literature values for the TB parameters of BLG.** The signs of the entries in table 1 marked by an asterisk have been corrected (using the table III in Jung and MacDonald[33]) to match the convention we used, which is the one from McCann and Koshino.[20] #Mallet *et al.* determined $v_F$ ($1.21 \times 10^6$ m/s) by determining directly the slope of the bands they measured; we translated this value for $v_F$ using $v_F = \sqrt{3}a\gamma_0/(2\hbar)$.

We compare our results to previously reported tight binding parameters for BLG in table 1. A comparison of our TB results to the values reported in the literature is intricate because of the disparity among definitions for TB parameters, which are not always explicit,[18,19,21–32] as discussed by Jung and MacDonald.[33] The signs of the entries in table 1 marked by an asterisk have been corrected following table III in Jung and MacDonald,[33] which matches the convention used by McCann and Koshino.[20] Our value of $\gamma_0$ is slightly higher than the other value reported for BLG on hBN.[21] However, compared to the work by Yankowitz *et al.*,[21] we probed the bands down to much larger binding energies and did not use an effective low energy Hamiltonian. ARPES analyses for BLG on SiC[30] and on SiO$_2$[19] led to substantially higher values for $\gamma_1$. The larger value found for BLG on SiC (0.48 eV)[30] can be explained by the greater interaction with the substrate in that case (evidenced by strong electron doping). The value of 0.61 eV found for BLG on SiO$_2$[19] is intriguing and might be explained by the difficulty in extracting $\gamma_1$ due to the observed low intensity of the top of the HEB.

We now discuss spatial mapping of our BLG/*h*BN heterostructure *via* nanoARPES. Figures 3b and 3c depict images constructed from a nanoARPES map of the BLG flake. At each point of this map, an ARPES E-k spectrum similar to the one displayed in Fig. 3a was recorded. During the map acquisition, the sample position was scanned in the xy-plane, while its rotational angles and the photoelectron analyzer were fixed. From this map, we constructed two different images (Fig. 3b and 3c) by integrating pixel intensity in the E-k regions outlined with brown (Fig. 3b) and blue (Fig. 3c) frames in Fig. 3a. Both images show dark spots, with close to zero intensity (two of which are indicated by white arrows in Figs. 3b and 3c). From a comparison between these ARPES images, atomic force microscopy (AFM) and optical images,[34] we conclude that these areas correspond to holes (absence of BLG) in the BLG flake. Such holes could have been created during the annealing process in UHV,[35] which is a required sample preparation step for nanoARPES experiments. Besides the shared presence of these dark spots on both images (Fig. 3b and 3c), there are also noticeable differences. Specifically, there is a greater intensity modulation in the region outlined by the white frame in Fig. 3c, when compared to the same region in Fig. 3b.

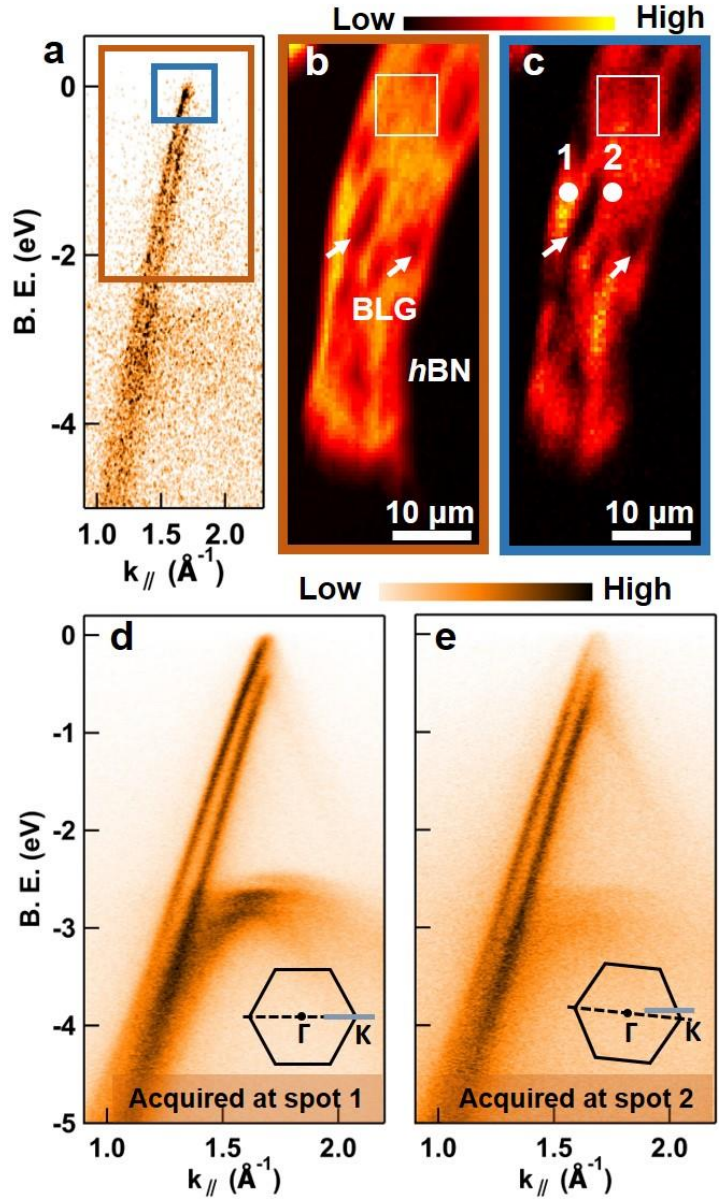

**Figure 3: Spatial variation of the BLG/*h*BN alignment evidenced with nanoARPES**. (a) Representative ARPES E-k spectrum from which the photoemission images shown in panels (b) and (c) are made. (b) Photoemission image of the BLG flake in which each pixel level is the integrated intensity within the brown rectangle depicted in (a). (c) Photoemission image of the BLG flake in which each pixel level is the integrated intensity within the blue rectangle depicted in (a). Arrows on (b) and (c) indicate areas where BLG is absent (holes). The region boxed by the white frames in both images illustrates the greater intensity modulation for image (c), compared to (b). (d) High resolution ARPES E-k spectrum obtained at point 1 on image (c) for which the analyzer entrance slit is parallel to the Γ-K direction, as indicated in the inset. (e) High resolution ARPES E-k spectrum obtained at point 2 on image (c), for which the analyzer entrance slit is slightly misaligned with the Γ-K direction, as indicated in the inset.

To gain further insight into this intensity fluctuation we acquired high resolution ARPES E-k spectra at various locations on the investigated sample. Fig. 3d and 3e show two typical ARPES E-k spectra obtained in regions labeled "1" and "2" and indicated by white dots in Fig. 3c. Notably, the intensity close to the Fermi level (B. E. =0 eV) is greater in Fig. 3d, compared to Fig. 3e. We attribute this difference to a variation of the BLG lattice alignment with the analyzer entrance slit, which is fixed. Indeed, the high intensity close to the Fermi level in Fig. 3d indicates that the analyzer entrance slit is aligned with the Γ-K direction for this spectrum, as indicated in the inset of Fig. 3d. This is not the case for the spectrum of Fig. 3e, where the loss of intensity at the Fermi level corresponds to a rotational misalignment between the analyzer entrance slit and Γ-K, as illustrated by the Brillouin zone schematics shown in the insets. To support this interpretation, we simulated ARPES E-k spectra for a BLG/analyzer misalignment of 0° and ~3° and found that the spectra of Figs. 3(d) and 3(e) were well reproduced.[34]

Using this insight, we now explain the intensity modulation differences between the images of Figs. 3b and 3c. The blue box in Fig. 3a (from which the image of Fig. 3c is made) corresponds to the very top of the LEB of BLG. Small misalignment between the probed BLG area and the electron analyzer decreases the integrated intensity in this area significantly as shown by the difference between Fig. 3d and 3e. On the contrary, the brown box in Fig. 3a is much larger and the corresponding integrated intensity is therefore less dependent on the lattice orientation of the probed BLG area. This explains the intensity modulation difference between the two images. Notably, because the supporting *h*BN flake is a single crystal, its orientation is the same throughout the entire device (*cf.* the STM analysis below). Thus, an alignment variation between the BLG and the electron analyzer equates to an alignment variation between the BLG and the *h*BN substrate.

To confirm the interpretation of our nanoARPES findings (Fig. 3) in terms of spatial variation of the BLG/*h*BN alignment, we acquired STM data on the same device that was studied in Figs. 1-3. Figures 4a and 4b show STM images obtained at two different locations on our sample and their respective fast Fourier transforms (FFTs) are presented in Fig. 4c and 4d. We observe moiré periodicities in the images of Fig. 4a and 4b of 11.6 nm and 4.1 nm, respectively. The relation between the moiré periodicity $\lambda$ and the misorientation angle $\varphi$ is $\lambda = (1 + \delta)a/\sqrt{2(1 + \delta)(1 - \cos \varphi) + \delta^2}$, where $a = 0.246$ nm is the graphene lattice constant and $\delta = 1.8\%$ is the lattice mismatch between graphene and *h*BN.[36] The observed moiré periodicities correspond to misorientation angles between the *h*BN and the BLG of 0.7° and 3.3°, respectively. We now discuss the FFTs of our data. The white (blue) dashed line in Fig. 4c (4d) indicates the BLG lattice orientation from the image in Fig. 4a (4b). In Fig. 4d we also reproduce the white dashed line from Fig. 4c to enable comparison between the lattice orientations. The 2.6° mismatch determined by the moiré wavelength comparison is confirmed by the direct comparison of the lattice orientation in the FFTs. Furthermore, we have observed intermediate periodicities at eight different locations on the sample such as 4.1 nm (misorientation of 3.3°), 4.4 nm (3.1°), 4.7 nm (2.8°), 5.5 nm (2.4°), 6.1 nm (2.1°), and 11.6 nm (0.7°). The observed variation in moiré periodicities firmly confirms our interpretation of the ARPES data in terms of BLG/*h*BN lattice alignment inhomogeneity.

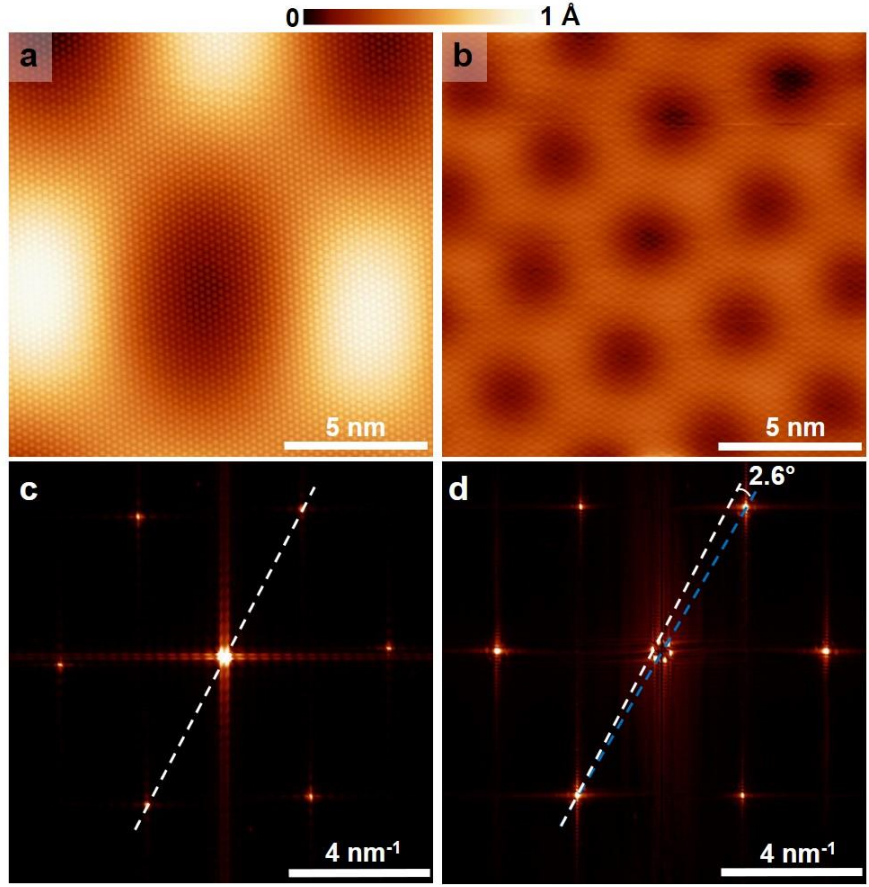

**Figure 4: Real-space STM imaging revealing different BLG/hBN alignment on the same device.** (a) and (b) are STM images obtained at two different locations on the same BLG/hBN heterostructure discussed in Figs. 1-3. The periodicities of the moiré patterns observed are 11.6 and 4.1 nm, respectively, corresponding to a BLG/hBN alignment angle of 0.7° and 3.3°. The tunneling current and sample bias used for (a) and (b) are 0.3 nA and 80 mV, and 0.1 nA and 75 mV, respectively. (c) and (d) are the fast Fourier transforms of images (a) and (b), respectively. The white dashed line on (c) serves as guide to the eye for the graphene lattice orientation of image (a) and is reproduced in (d). The blue dashed line in (d) indicates the orientation of the graphene lattice on image (b). The angle between these two lines is 2.6°, as expected from the moiré wavelengths.

We believe that the rotational disorder seen in our experiment appears during the annealing procedure because of the heat-induced self-alignment mechanism first reported by Woods *et al.*[37] These authors have shown that it is energetically favorable for graphene to be macroscopically aligned within +/- 0.7° with *h*BN, and that mild heating (200°C in Ar/H$_2$ environment) can provide the energy necessary for self-alignment of regions that are misaligned by small angles (close to 1°). In particular, they demonstrated quasi homogeneous rotation by ~0.3° for flakes with characteristic size of ~30 µm. In their experiment, they did not observe self-alignment for samples displaying more than a few bubbles. This was attributed to the contaminants within bubbles decoupling graphene from $h$BN or acting as pinning centers.[37] In our experiment, we believe the misalignment between the BLG and the $h$BN was ~3.3° after transfer and was reduced in a spatially inhomogeneous fashion by the UHV annealing, down to 0.7° at certain areas. The self-rotation we observed (>2.5°) is thus much larger than what was reported by Woods *et al*. and is inhomogeneous across our sample. We attribute this inhomogeneity to the fact that our sample contained numerous bubbles that burst during the UHV annealing.[34] This most likely segmented our flake into smaller domains loosely attached to one another and thus able to rotate more easily.

In conclusion, we have presented the results from a nanoARPES investigation of a BLG/$h$BN heterostructure. We have directly extracted TB parameters with nanoARPES for BLG on $h$BN, the standard insulating supporting substrate for 2D material heterostructures. We have also shown that nanoARPES can reveal variations in lattice alignment between BLG and $h$BN. These latter findings were confirmed by STM imaging, which showed areas with different moiré periodicities due to numerous domains with various orientations. We attribute this variation of lattice alignment to an inhomogeneous BLG/$h$BN alignment induced by standard UHV heating used to prepare samples for ARPES and STM experiments. These results substantially improve our understanding of the BLG/$h$BN heterostructure and provide researchers using this heterostructure with a direct picture of the electronic bands resolved in reciprocal space with nanoscale resolution. In addition, the spatial variation of lattice alignment between the BLG and the supporting $h$BN we evidenced here is important for researchers investigating these samples with spatially averaging probes.

We thank Jeil Jung for useful discussions. K.W. and T.T. acknowledge support from the Elemental Strategy Initiative conducted by the MEXT, Japan and the CREST (JPMJCR15F3), JST. F.J., J.A., M.C.A., and J.V.J. conceived the work and designed the research strategy. E.Q., J.L.D., H.C., and F.J. fabricated the sample, under J.V.J.'s supervision. K.W. and T.T. provided the $h$BN crystals. J.A., F.J., and C.C. performed the nanoARPES experiments, under M.C.A.'s supervision. F.J. and E.Q. performed the STM experiments, under J.V.J.'s supervision. E.Q. performed the AFM measurements. F.J. analyzed the data. F.J. and J.V.J. wrote the paper. All authors discussed the paper and commented on the manuscript.